\documentclass[reprint,aps,superscriptaddress,prl]{revtex4-2}
\usepackage[table]{xcolor}
\usepackage[linktocpage=true]{hyperref}
\hypersetup{colorlinks=true,citecolor=red,urlcolor=blue}
\usepackage{amsmath}
\usepackage{amssymb}
\usepackage{braket}
\usepackage{graphicx}
\usepackage{dcolumn}
\usepackage{bm}
\usepackage{gensymb}
\usepackage{multirow}
\usepackage{changes}
\usepackage{xcolor,soul}
\sethlcolor{cyan}
\begin{document}
\preprint{}
\title{Multi-photon Atom Interferometry via cavity-enhanced Bragg Diffraction}
\author{D. O. Sabulsky}
\author{J. Junca}
\author{X. Zou}
\author{A. Bertoldi}
\affiliation{LP2N, Laboratoire Photonique, Num{\'e}rique et Nanosciences, Universit{\'e} Bordeaux--IOGS--CNRS:UMR 5298, rue F. Mitterrand, F--33400 Talence, France}
\author{M. Prevedelli}
\affiliation{Dipartimento di Fisica e Astronomia, Universit\`{a} di Bologna, Via Berti-Pichat 6/2, I-40126 Bologna, Italy}
\author{Q. Beaufils}
\author{R. Geiger}
\author{A. Landragin}
\affiliation{LNE--SYRTE, Observatoire de Paris, Universit{\'e} PSL, CNRS, Sorbonne Universit{\'e}, 61 avenue de l'Observatoire, F--75014 Paris, France}
\author{P. Bouyer}
\author{B. Canuel}
   \email{benjamin.canuel@institutoptique.fr}
   \affiliation{LP2N, Laboratoire Photonique, Num{\'e}rique et Nanosciences, Universit{\'e} Bordeaux--IOGS--CNRS:UMR 5298, rue F. Mitterrand, F--33400 Talence, France}
   \collaboration{MIGA Consortium}
\date{\today} 
\begin{abstract}
We present a novel atom interferometer configuration that combines large momentum transfer with the enhancement of an optical resonator for the purpose of measuring gravitational strain in the horizontal directions. 
Using Bragg diffraction and taking advantage of the optical gain provided by the resonator, we achieve momentum transfer up to $8\hbar k$ with mW level optical power in a cm-sized resonating waist.  
Importantly, our experiment uses an original resonator design that allows for a large resonating beam waist and eliminates the need to trap atoms in cavity modes. 
We demonstrate inertial sensitivity in the horizontal direction by measuring the change in tilt of our resonator.
This result paves the way for future hybrid atom/optical gravitational wave detectors. 
Furthermore, the versatility of our method extends to a wide range of measurement geometries and atomic sources, opening up new avenues for the realization of highly sensitive inertial atom sensors.
\end{abstract}
\maketitle
Improving the sensitivity and accuracy of matter-wave based quantum sensors beyond the current state-of-the-art has garnered significant interest due to its wide range of applications, ranging from terrestrial and space-based gravitational wave observation \cite{Dimopoulos2009,Graham2013,Bertoldi2021_1,Canuel2018, Canuel_2020, Badurina_2020, Schlippert2020, Abe2021}, studies of dark matter \cite{Arvanitaki2018,El-Neaj2020, PhysRevD.105.023006} or tests of general relativity \cite{PhysRevLett.120.183604, PhysRevLett.125.191101,Overstreet2022}, to geophysics \cite{Canuel2016,AntoniMicollier2022,Stray2022} or geodesy \cite{Trimeche2019}.
Reaching these goals requires multiple methods, which can be combined in a synergistic fashion; these methods include enhancing the atomic flux \cite{Loriani2019}, implementing interleaved interferometers \cite{Savoie2018}, employing entangled sources \cite{PhysRevLett.127.140402,Greve2022} and large momentum transfer (LMT) atom optics \cite{McGuirk2000,Cadoret2008,Muller2008_1,Muller2009PRL,Kovachy2015,PhysRevLett.116.173601,Jaffe2018,Gebbe2021,PhysRevA.102.053312,PhysRevLett.129.183202}.
\begin{figure}[ht!]
\centering
\includegraphics[width=0.9\linewidth]{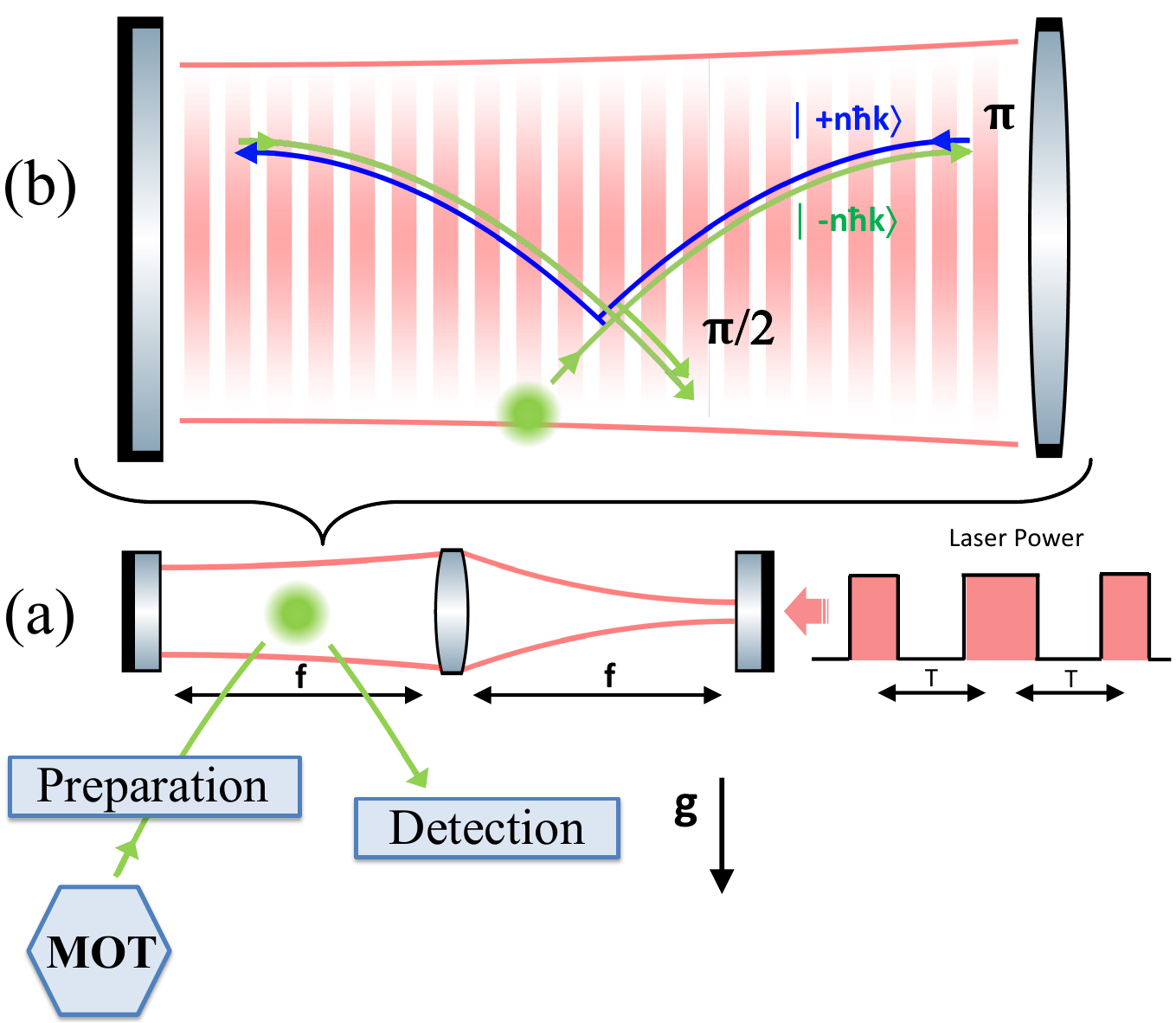}
\caption{(a) Marginally stable resonator geometry comprising two plane mirrors, positioned at the focal plane of an intra-cavity lens. 
The atoms undergo a sequence of operations including trapping, cooling, vertical launch, state selection, and velocity filtering prior to injection into the semi-degenerate mode volume of the horizontal resonator, where Bragg diffraction is driven via time-modulation of the light injected into the resonator.
After interrogation, the transition probability is measured using a momentum-selective detection.
(b) Geometry of the $\pi/2$ - $\pi$ - $\pi/2$ Mach-Zehnder type atom interferometer utilizing Bragg diffraction for manipulating the momentum states $\left |\pm n\hbar \mathbf{k}\right\rangle$. 
\label{fig:0}}
\end{figure}
\par Enhancing and controlling the atom-light interaction is a fundamental prerequisite for these methods, and optical cavities offer exceptional capabilities in this regard.
They provide an ideal platform for trapping and cooling atoms within their modes \cite{Naik_2018} - \textcolor{black}{previous demonstrations of atom interferometry in an optical cavity were limited to the realization of a standard two photon cavity beamsplitter \cite{Hamilton2015}.}
Leveraging the enhanced coupling between the atoms and the cavity mode allows for improved detection sensitivity, surpassing the limitations imposed by shot noise \cite{Greve2022}.
In addition, optical resonators offer the potential for realizing enhanced beam splitters by effectively cleaning the spatial mode of the interrogation beam and amplifying its intensity. 
This feature is particularly crucial for LMT methods that rely on Bragg diffraction or Bloch oscillations, as these techniques impose stringent laser power requirements.
\textcolor{black}{Importantly, these demonstrations rely upon trapping and manipulating atomic ensembles within the mode volume of the cavity, which is oriented parallel with gravity - this \textit{strongly restricts} the applications, placing narrow constraints on the injected momentum distribution of the ensemble, but also for the physics that can be explored, in particular in view of gravitational wave detection which requires long baseline horizontal resonators.}
\begin{figure*}[t]
\centering
\includegraphics[width=1\linewidth]{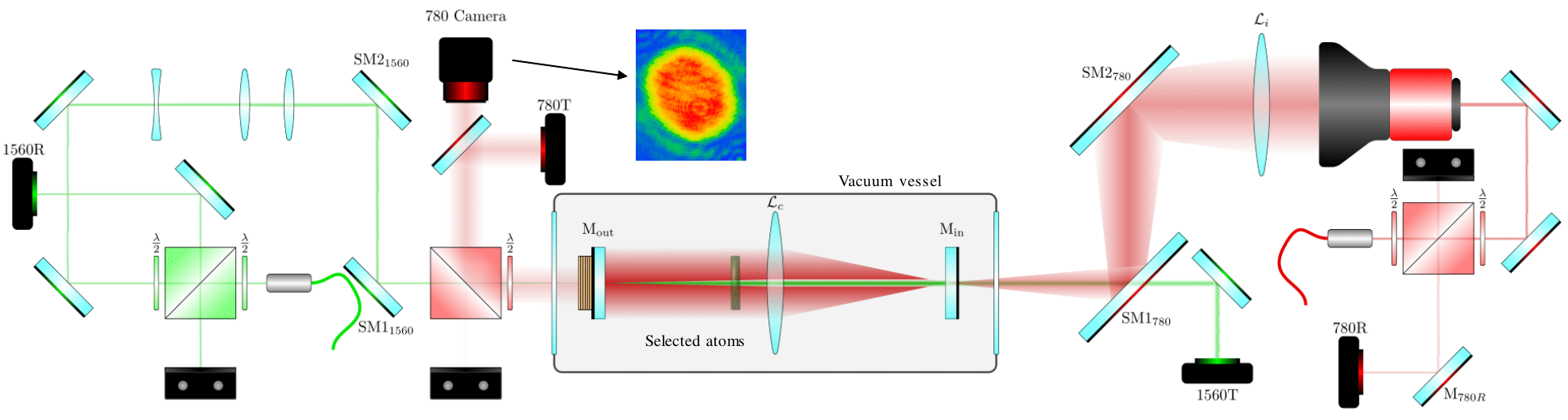}
\caption{Schematic of the laser system for injecting the marginally stable optical resonator.
Gaussian beams at 1560 nm (green path) and 780 nm (red path) are injected into opposite sides of resonator. 
These beams are inherently phase-locked, originating from the same source.
To facilitate injection, the beams pass through their respective telescopes and are focused onto the mirrors.
The 1560 nm beam is employed for length-frequency stabilization of the resonator with a piezostack on $\text{M}_{\text{out}}$.
\label{fig:1}} 
\end{figure*}
Moreover, while the influence of cavity response on the shape of the light pulses used to coherently manipulate the matter waves \cite{Fang2018} has been traditionally considered a limitation of cavity interrogation \cite{Dovale2017}, recent advancements have introduced innovative approaches to overcome this challenge. 
These include the use of light-shift engineering \cite{Bertoldi2021} and intracavity frequency modulation of circulating pulses \cite{Nourshargh2021,Nourshargh:22}, enabling the utilization of high finesse or long baseline interrogation resonators. 
These developments open up new possibilities for cavity-based atom interferometry, facilitating the realization of highly sensitive quantum sensors and hybrid atom-optic gravitational wave detectors based on long optical cavities. 
This novel generation of instruments would leverage the simultaneous utilization of atom and optical readouts to provide an expanded gravitational wave detection band.
\par In this letter, we study and demonstrate this new atom interferometry configuration using enhanced Bragg diffraction with momentum transfer up to 8 $\hbar$k in combination with an original configuration of a \textit{horizontal} optical resonator.
Our experiment takes advantage of a specific resonator design where we achieve large waist resonating modes allowing for the atoms do not to be trapped in the cavity mode. 
The cavity is a marginally stable resonator \cite{Riou2017, Mielec:20}, as illustrated in Fig.~\ref{fig:0}. 
It consists of two high-reflectivity mirrors placed at the focal planes of a lens, which allows for the propagation of large and arbitrary spatial modes \textcolor{black}{and significantly lowers the required optical power to drive Bragg diffraction}.
We achieve a resonating mode with a 4 mm FWHM in an 80 cm long horizontal resonator.
Atoms launched vertically interact with the resonating field when reaching the apogee of their parabolic trajectory. 
By pulsing the laser power injected into the resonator, we realize a three pulse $\pi/2$ - $\pi$ - $\pi/2$ Mach-Zehnder type matter wave interferometer operating in the quasi-Bragg regime. 
We achieve a momentum transfer of up to $2n = 8 \hbar k$ and demonstrate inertial sensitivity to horizontal accelerations by monitoring the tilt of the entire experiment. 
\textcolor{black}{In a context where optical cavities have long been considered an improvement channel for matter-wave interferometers, we thus demonstrate, for the first time, the ability to overcome standard 2-photon atom interferometry using a cavity beamsplitter.}
\textcolor{black}{This configuration, wherein the atomic ensemble is not trapped in the cavity mode volume, opens the applicability of this method to a vast class of measurement geometries.}
\begin{figure*}[]
\centering
\includegraphics[width=1\linewidth]{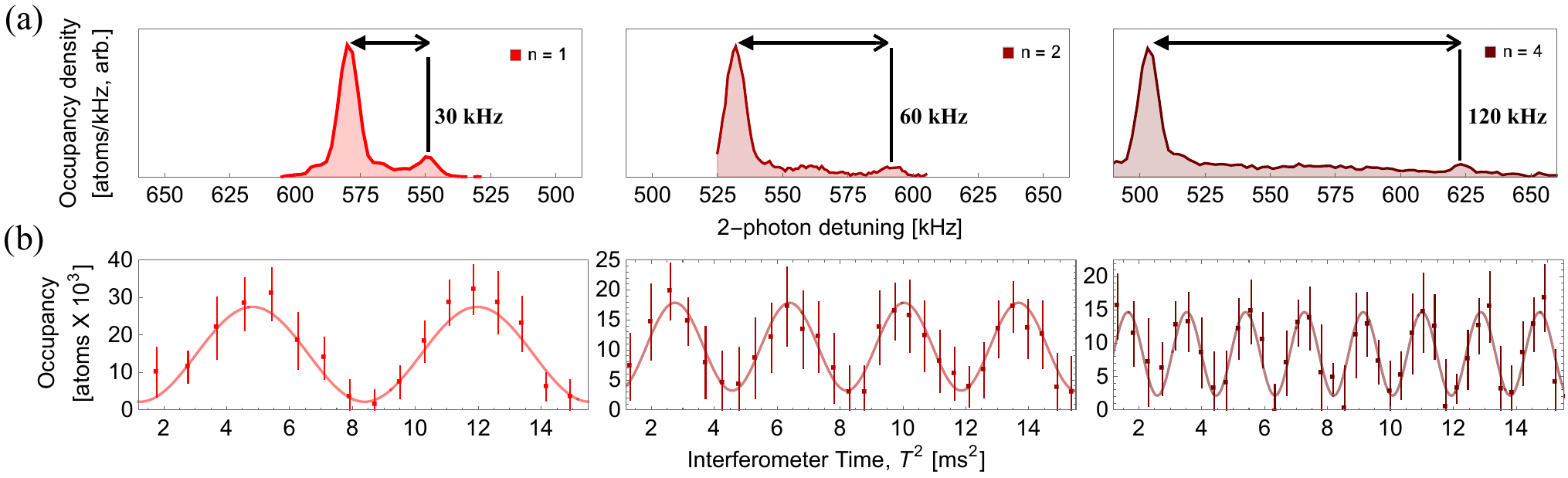}
\caption{Characterization of Bragg diffraction and interferometer performance. 
(a) Raman spectroscopy with a $\pi$-pulse, revealing the atomic momentum distribution following a $\pi$-pulse of the Bragg diffraction process.
The data correspond to $n=1,2,4$, displayed from left to right. 
Each data point represents an average of 10 measurements. 
(b) Number of atoms in the diffracted state as a function of the interferometer time for $n = 1, 2, 4$, presented from left to right. 
Each data point represents an average of 20 measurements. 
The dashed lines depict fits of the fringe patterns. 
The tilt angle remained constant during these measurements. 
\label{fig:3}}
\end{figure*}
\par The laser system used for Bragg diffraction has unique features that are shown in Fig.~\ref{fig:1}.
The system is based on a fibered laser diode centered at 1560 nm with a Lorentzian FWHM of 4.30 kHz.
This laser serves as the seed for a high-power dual-stage Erbium-doped fiber amplifier, which, in turn, pumps a periodically-poled Lithium-Niobate crystal.  All beams are shunted over polarization-maintaining fibers, to finally generate the diffraction light at 780 nm. 
The 1560 nm and 780 nm lasers are injected into opposite ports of the optical resonator, after passing through the viewports of a vacuum chamber that maintains the resonator at a residual pressure $<10^{-10}$ mbar.
Precise alignment of the optical  is achieved using UHV-compatible linear piezo actuators, with a step resolution of $<30$ nm, housed within kinematic mirror mounts.
To ensure complete control, we employ three actuators on the injection mirror for 780 nm beam and two actuators along with a bored cylindrical piezoelectric transducer stack on the reflecting mirror.
\par The 1560 nm diode laser is servoed to the resonator using phase modulation via a fibered EOM, employing the Pound-Drever-Hall technique \cite{Drever1983}. Fast feedback is applied to the laser diode current, and to prevent large DC actuation of this current, slow feedback is applied to the  piezostack (PZT).
The frequency doubled light at 780 nm is modulated using a fibered AOM to generate the diffraction pulses. 
We employ Gaussian amplitude modulation of a sinusoidal carrier produced by an arbitrary function generator. 
It is important to note that in this laser architecture, the diffraction and resonator servo frequencies are naturally phase coherent, making it significantly simpler compared to previously reported systems for stable cavities \cite{Hamilton2015}. 
Furthermore, this architecture is highly adaptable to different resonating mode sizes, offering flexibility in experimental setups.
\par A detailed description of our optical resonator and its limitation can be found in our previous works \cite{Riou2017, Mielec:20}. The  optics consist of two 25 mm flat mirrors with 99,7\% reflectivity, featuring peak-to-valley surface irregularities of $\lambda/55$ and $\lambda/73$, respectively. Additionally, a 50 mm bi-convex lens with a focal length of 400 mm, exhibiting peak-to-valley errors of $\lambda/5$ and $\lambda/3$ (measured at 633 nm) is included. The choice of a bi-convex lens over a plano-convex lens ensures injection symmetry for laser-cavity stabilization and avoids spurious cavities. 
To generate the circulating mode, we inject a Gaussian beam focused to a spot characterized by a $1/e^{2}$ diameter of $\approx$ 30 $\mu$m on the back of the first  mirror.
With this setup, we measure an optical gain of $G=38$, as effective finesse of $\mathcal{F}=200$ and a maximum resonating beam diameter about 4 mm FWHM. 
Our previous investigations \cite{Riou2017,Mielec:20} have revealed that the mode diameter is limited by the optical aberrations of the setup, particularly the astigmatism and spherical aberrations originating from the intra-cavity lens.
The achieved optical gain is sufficient to drive diffraction with mW level input from a low-noise source, while the finesse ensures that the temporal response of 0.17 $\mu$s is significantly smaller than the Bragg diffraction pulse, $\tau \sim \mathcal{O}$(100 $\mu$s), and so does not distort the Gaussian shape.
\par We launch a dilute atomic ensemble with an initial vertical velocity of $v \sim 3.8$ m/s into the resonator mode volume. 
The ensemble, composed of $10^6$ $^{87}$Rb atoms at a temperature of $2.3$ $\mu$K \cite{Sabulsky2020}, passes through a preparation region (see Fig~\ref{fig:0} and \cite{Beaufils2022}).
The atoms are prepared in the $F = 1$ hyperfine ground state with $m_{F}=0$. 
To ensure optimal conditions for the Bragg diffraction, a stringent velocity pre-selection is applied to the ensemble along their ballistic trajectory, resulting in a narrow momentum distribution of 0.2 $v_{\text{rec}}$ HWHM along the diffraction axis. 
The preparation protocol reduces the number of atoms on the original ballistic trajectory towards the resonator to a few $10^{5}$.
A narrow momentum distribution facilitates single frequency quasi-Bragg diffraction, as efficient and monochromatic pulses typically require tens to hundreds of microseconds.
Additionally, it enables precise control of the insertion angle with respect to the geometric axis of the mode volume, leading to higher contrast in the resulting interferometer. 
The ballistic trajectory of the ensemble intersects the axis of the resonator at its apogee, where we apply Bragg diffraction pulses.
\par The Bragg laser system is blue-detuned by $\Delta=3.4$ GHz relative to the $F = 1 \longrightarrow F' = 2$ transition in $^{87}$Rb; this is the so-called ``quasi-Bragg" scattering regime characterized by short interaction times and reduced losses to intermediate momentum states.
We apply the diffraction pulses with a Gaussian time envelope, which has been demonstrated to enhance the efficiency and monochromaticity of Bragg diffraction \cite{Muller2008a}.
After the completion the Bragg interferometer, the ensemble continues its ballistic trajectory back into the preparation region. 
Here, we employ a state labeling detection system \cite{Beaufils2022,PhysRevA.98.043611} based on Doppler sensitive Raman spectroscopy.
This operation is followed by fluorescence detection of the atomic population in each hyperfine state using resonant light sheets.
\begin{figure*}[]
\centering
\includegraphics[width=1\linewidth]{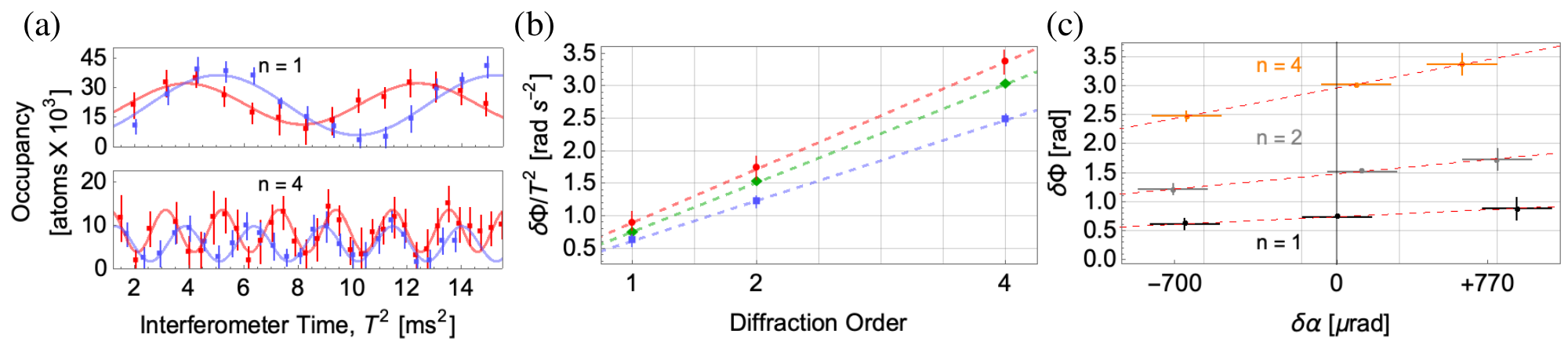}
\caption{ Demonstrating inertial sensitivity. 
(a) Fringe pattern obtained from three pulse Bragg interferometers by scanning the dwell time $T$ of the interferometer. 
Results are shown for $n = 1$ and $4$ using two different angles $\alpha$ (red and blue). 
Each data point represents an average of 20 measurements.
(b) Scale factor of the atom interferometer as a function of the diffraction order for three different tilts $\alpha$, normalized to $T = 1$ second. 
The fringe patterns from a) correspond to the points for $n = 1$ and $4$, in red and blue.  
\textcolor{black}{We applied different tilts to the experiment, extracted from the fits as $2203\pm8$ $\mu$rad (red), $1378\pm8$ $\mu$rad (green), and $620\pm7$ $\mu$rad (blue).}
%$826\pm8$ $\mu$rad (red), $758\pm8$ $\mu$rad (green), and $620\pm7$ $\mu$rad (blue)
(c) Inertial sensitivity of the atom interferometer as a function of the tilt angle, separated by diffraction order, normalized to $T = 1$ second. 
The dashed red lines are calculated $\delta\Phi$ for each diffraction order $n$.
\label{fig:4}}
\end{figure*}
\par In Fig.~\ref{fig:3}, we demonstrate multi-photon Bragg interferometry using the optical resonator described here and the atomic source described in \cite{Beaufils2022}.
We first observe Bragg diffraction within the resonator via Raman spectroscopy with a $\pi$-pulse, driving up to four diffraction orders with a beam diameter of 4 mm, as depicted in panel (a) of Fig.~\ref{fig:3}.
We observe minor spurious excitation beyond the targeted momentum state. 
The transfer efficiency decreases with increasing diffraction order, from $7.7\%$ for $n = 1$, to $4.0\%$ for $n = 2$, and further down to $3.6\%$ for $n = 4$ \textcolor{black}{- the efficiency is determined by comparison to the original velocity class, the large peak in all orders of (a) in Fig.~\ref{fig:3}. }
These low transfer efficiencies are a matter of technical concern - the unexpectedly low optical quality of the biconvex lens limits the effective beam diameter while the problem is exacerbated by the approximately 1 cm diameter of the ensemble at the interaction region.
For driving a Bragg diffraction $\pi$-pulse, the injected optical power into the resonator varies from 900 $\mu$W for $n = 1$, to 3 mW for $n = 2$, and finally up to 18 mW for $n = 4$.
We use a $\pi$-pulse duration corresponding to the Gaussian FWHM values of 102 $\mu$s, 51 $\mu$s, and 32 $\mu$s for $n = 1, 2, 4$, respectively, while maintaining the same beam size. 
We then apply a $\pi/2$ - $\pi$ - $\pi/2$ sequence of  pulses and observe interference patterns by varying the dwell time  $T$ between pulses. 
\par The largest free fall distance between the first and last pulse in the sequence is on the order of a few millimeters.
In this configuration, the axis of Bragg diffraction, which is aligned with the geometric axis of the resonator, has a small angle with the horizontal axis, leading to a measurable acceleration from a residual angle with gravity. 
We scan the interferometer time $T$, thereby sweeping the scale factor, spanning this residual acceleration due to gravity. 
To characterize the interference pattern, we present the fringes in terms of atom number in the targeted momentum state, which we refer to as occupancy - the atoms that actively participated in the interferometer. 
In (b) of Fig.~\ref{fig:3}, we display the fringes for different target momentum states $n = 1, 2, 4$, presented from left to right. 
The decrease in the number of participating atoms for increasing $n$ is expected \cite{Riou2017,Mielec:20}. 
In the quasi-Bragg regime, the transfer of atoms to the target momentum state is a non-linear process: as the effective Rabi frequency $\Omega_{\text{eff}}$ increases \cite{Muller2008_1,Riou2017}, the process becomes highly sensitive to transverse intensity inhomogeneity.
\par To demonstrate the inertial sensitivity of the interferometer we intentionally introduce a tilt to the entire instrument, thereby modifying the angle $\alpha$ between the Bragg diffraction axis and the axis perpendicular to Earth's gravity - prior work on atom interferometer tiltmeters and quantum inertial sensor bench-marking using tilts with gravity pave the way for this demonstration \cite{PhysRevLett.116.173601,Dutta2016,PhysRevA.100.053618,PhysRevA.105.013316}.
We quantify the tilt variation, denoted $\delta\alpha$, through two different measurements: velocity-selective Raman spectroscopy, \cite{Junca:2022nwo} and (a) of Fig.~\ref{fig:3}, and Bragg interferometry. 
We apply a sequence of Bragg pulses ($\pi/2$ - $\pi$ - $\pi/2$) to the resonator, with varying time intervals $T$. 
We measure the occupancy of the resulting ensemble for the two different instrument orientations, as depicted in panel (a) of Fig.~\ref{fig:4}. 
The tilt in the experiment alters the projection of the local gravity $g$ along the measurement axis, resulting in an atomic phase shift $\Phi=2 n \alpha g k T^2$. 
From the fringe pattern in panel (a) of Fig.~\ref{fig:4}, we extract the scale factor $\Phi/T^2$, which is plotted in panel (b) of the same figure for the different orientations and diffraction orders.
For a given diffraction order, the shift in the scale factor between the two orientations is related to $\delta\alpha$ through the equation: $\delta(\Phi/T^2)=2 n \delta\alpha g k$.
The data presented in (b) of Fig.~\ref{fig:4} shows a linear increase in the measurement scale factor as a function of $n$, which is consistent with the expected behavior for an interferometer employing LMT. 
This behaviour has been verified for various orientations and resonating beam sizes within the device, demonstrating that the marginally stable resonator does not introduce optical limitations for manipulating momentum states within an atom interferometer.
Finally, in (c) of Fig.~\ref{fig:4} and normalized for $T = 1$ s, we show the change in total phase $\delta\Phi$ as a function of the change in tilt $\delta\alpha$ and find good agreement with the predicted curves as a function of diffraction order. 
\par In this letter, we have demonstrated multi-photon atom interferometry inside an optical resonator and established inertial sensitivity of the device through tilt experiments.
This work marks a significant step towards enhancing the sensitivity of matter wave interferometry experiments. 
In particular, atom interferometry is gaining recognition as a potential approach for constructing large-scale detectors to observe gravitational waves and investigate dark matter \cite{Zhan2019,Canuel_2020,Badurina_2020,Abe2021,Alonso2022}. 
These experiments would require interferometer pulses spanning tens of kilometer in the horizontal direction, and the resonators presented here could play a pivotal role in realizing such systems \cite{Dovale2017}.  
A key aspect of our demonstration is the utilization of a marginally stable resonator to enhance the light-matter interaction. 
This approach enables a flexible degenerate resonating mode size with no mode structure, allowing the entire cavity volume to be used for driving diffraction.
In contrast, stable cavities restrict efficient interaction to small mode volumes. 
As a result, this method hold promise for diverse atom interferometer sources and geometries, including mobile, sub-Doppler cooled vertical gravimeters \cite{Louchet_Chauvet_2011}, horizontal gyroscopes \cite{Dutta2016}, and mobile rotation sensors and clocks attempting employing spin-squeezing \cite{PRXQuantum.4.020322}. 
%
%
%
%\newpage
\begin{acknowledgments}
The authors would like thank I. Riou, G. Lef\`{e}vre, and N. Mielec for their early work on the experiment.
We gratefully acknowledge the engineering support of P. Teulat and L. Sidorenkov.
J.J. thanks the ``Association Nationale de la Recherche et de la Technologie'' for financial support (N$^\mathrm{o}$ 2018/1565). 
X.Z. thanks the China Scholarships Council (N$^\mathrm{o}$ 201806010364) program for financial support.
This work was realized with the financial support of the French State through the ``Agence Nationale de la Recherche'' (ANR) within the framework of the ``Investissement d'Avenir'' programs Equipex MIGA (ANR-11-EQPX-0028) and IdEx Bordeaux - LAPHIA (ANR-10-IDEX-03-02). 
This work was also supported by the r{\'e}gion d'Aquitaine (project USOFF), and the European Union’s Horizon 2020 research and innovation programme (Grant CRYST$^3$ No. 964531). 
We also acknowledge support from the CPER LSBB2020 project; funded by the ``r{\'e}gion PACA'', the ``d{\'e}partement du Vaucluse'', and the ``FEDER PA0000321 programmation 2014-2020''. 
And finally, we acknowledge financial support from Ville de Paris (project HSENS-MWGRAV) and Agence Nationale pour la Recherche (PIMAI ANR-18-CE47-0002-01 and EOSBECMR ANR-18-CE91-0003-01).
\end{acknowledgments}

\bibliography{apsbib.bib}

\end{document}